\documentclass[twocolumn,showpacs,aps,prb,amssymb,amsmath]{revtex4}
\usepackage{amsmath}
\usepackage{graphicx}
\usepackage{amssymb}
\usepackage{amsfonts}

\begin{document}

\title{Non-ohmic critical fluctuation conductivity of layered superconductors
in magnetic field}

\author{I. Puica}

\email{ipuica@ap.univie.ac.at}

\author{W. Lang}

\affiliation{Institut f\"{u}r Materialphysik der Universit\"{a}t Wien, Boltzmanngasse
5, A-1090 Wien, Austria}

\begin{abstract}
Thermal fluctuation conductivity for a layered superconductor in perpendicular
magnetic field is treated in the frame of the self-consistent Hartree approximation
for an arbitrarily strong in-plane electric field. The simultaneous application
of the two fields results in a slightly stronger suppression of the superconducting
fluctuations, compared to the case when the fields are applied individually.
\end{abstract}


\maketitle
The high-temperature superconductors (HTSC) show a more pronounced effect
of fluctuations in the normal-superconducting transition region, due to
their high critical temperature, small coherence length, and quasi-two-dimensional
nature. Outside the critical region above $T_{c}$, in the absence of magnetic
field and for small electric fields, the fluctuation conductivity can be
explained by the Aslamazov-Larkin\cite{Aslamazov68} theory, subsequently
extended by Lawrence and Doniach (LD)\cite{Lawrence71} for two-dimensional
layered superconductors, a situation very much resembling the crystal structure
in cuprates.

The fluctuation conductivity may be however calculated in the linear-response
approximation only for sufficiently weak electric fields, that do not perturb
the fluctuation spectrum.\cite{Hurault69} Reasonably high values of the
electric field accelerate the fluctuating paired carriers to the depairing
current, and suppress the fluctuation lifetime, leading to deviation from
the Ohm's law. In connection with the low-temperature superconductors,
the nonlinearity has been studied theoretically for the isotropic case,\cite{Schmid69,Tsuzuki70}
while more recently Varlamov and Reggiani\cite{Varlamov92} and Mishonov
\emph{et al.}\cite{Mishonov02} addressed the issue of the non-ohmic fluctuation
conductivity for a layered superconductor in an arbitrary electric field.
The above mentioned theories describe however the fluctuations as non-interacting
Gaussian ones, neglecting the quartic term in the Ginzburg-Landau (GL)
free energy. This approximation holds not too close to the transition point,
but it breaks down in the critical region, for higher densities of fluctuation
Cooper pairs.

The Ohmic fluctuation conductivity in the presence of a magnetic field
was also initially treated\cite{Maki69,Klemm74} in the Gaussian fluctuation
approach, which predicted a divergence at $T_{c}(H)$ that is however not
observed. The physical reason is the motion of vortices providing dissipation
and hence a finite flux-flow conductivity. Ikeda \emph{et al.}\cite{Ikeda91a}
and Ullah and Dorsey (UD)\cite{Ullah91} showed that the theoretical divergence
can be eliminated by using the Hartree approximation, which treats self-consistently
the quartic term in the GL free-energy expansion. This approach was applied
for the longitudinal\cite{Ikeda91a,Ullah91} and Hall conductivity,\cite{Ullah91}
for the Nernst effect and the thermopower,\cite{Ullah91} in the linear-response
approximation, for a layered superconductor under magnetic field.

In this paper we shall treat, in the self-consistent Hartree approximation,
the thermal fluctuation conductivity for a layered superconductor in a
perpendicular magnetic field and for an arbitrarily strong in-plane electric
field, a topic that, to our present knowledge, has not been treated yet.
While the effect of the interacting fluctuations under applied magnetic
field was investigated in the Hartree model\cite{Ikeda91a,Ullah91} only
in the linear response approximation, that is for infinitesimally small
electric fields, the non-linear conductivity for a layered system under
arbitrarily strong electric field was derived in the Gaussian,\cite{Varlamov92,Mishonov02}
as well as in the Hartree approximation,\cite{PuicaLangE} but only in
the absence of magnetic field.

For our purpose, we shall adopt the Langevin approach to the time-dependent
Ginzburg-Landau (TDGL) equation.\cite{Schmid69,Ullah91} The starting point
of the formalism is the same as in Refs. \onlinecite
{Ullah91} and \onlinecite
{PuicaLangE}, and consists of the LD expression of the GL free energy for a system
of Josephson coupled superconducting planes, the gauge-invariant relaxational
TDGL equation governing the critical dynamics of the superconducting order
parameter, and the Langevin white-noise forces that model the thermodynamical
fluctuations, satisfying the fluctuation-dissipation theorem. Keeping the
same notations as in Ref. \onlinecite
{PuicaLangE}, the TDGL equation for the superconducting order parameter in the $l$-th
plane will write\begin{align}
\Gamma _{0}^{-1}\frac{\partial \psi _{l}}{\partial t}-i\frac{e_{0}\Gamma _{0}^{-1}Ex}{\hbar }\psi _{l}+a\psi _{l}+b\left|\psi _{l}\right|^{2}\psi _{l} & \nonumber \\
-\frac{\hbar ^{2}}{2m}\left[\partial _{x}^{2}+\left(\partial _{y}-\frac{i\, e_{0}}{\hbar }xB\right)^{2}\right]\psi _{l} & \nonumber \\
+\frac{\hbar ^{2}}{2m_{c}s^{2}}(2\psi _{l}-\psi _{l+1}-\psi _{l-1})=\zeta _{l}\left(\mathbf{x},t\right)\; . & \label{EQini}
\end{align}
 Here $m$ and $m_{c}$ are effective Cooper pair masses in the $ab$-plane
and along the $c$-axis, respectively, $s$ is the distance between superconducting
planes, and the pair electric charge is $e_{0}=2e$. The order parameter
has the same physical dimension as in the 3D case, and SI units are used.
The perpendicular magnetic field $B$ is generated by the vector potential
in the Landau gauge, $\mathbf{A}=\left(0,xB,0\right)$, with $x$ and $y$
the in-plane coordinates, and the magnetization is neglected. The GL potential
$a=a_{0}\varepsilon $ is parameterized by $a_{0}=\hbar ^{2}/2m\xi _{0}^{2}=\hbar ^{2}/2m_{c}\xi _{0c}^{2}$
and $\varepsilon =\ln \left(T/T_{0}\right)$, with $T_{0}$ the mean-field
transition temperature, while $\xi _{0}$ and $\xi _{0c}$ are, respectively,
the in-plane and out-of-plane coherence lengths extrapolated at $T=0$.
The relaxation rate $\Gamma _{0}$ in the TDGL equation is given by\cite{Masker69}
$\Gamma _{0}^{-1}=\pi \hbar ^{3}/16m\xi _{0}^{2}k_{B}T$. The Langevin
white-noise forces $\zeta _{l}\left(\mathbf{x},t\right)$ are correlated
through $\left\langle \zeta _{l}\left(\mathbf{x},t\right)\zeta _{l'}^{*}\left(\mathbf{x}',t'\right)\right\rangle =2\Gamma _{0}^{-1}k_{B}T\delta (\mathbf{x}-\mathbf{x}')\delta (t-t')\delta _{ll'}/s$,
where $\delta (\mathbf{x}-\mathbf{x}')$ is the 2-dimensional delta-function
concerning the in-plane coordinates. The electric field $\mathbf{E}$ is
assumed along the $x$-axis, generated by the scalar potential $\varphi =-Ex$.
In the chosen gauge, the current density operator along the $x$ direction
in the $l$-th plane will give, after averaging with respect to the noise\begin{equation}
\left\langle j_{x}^{(l)}\right\rangle =\left.-\frac{ie_{0}\hbar }{2m}(\partial _{x}-\partial _{x^{\prime }})\left\langle \psi _{l}\left(\mathbf{x},t\right)\psi _{l}^{*}\left(\mathbf{x}',t\right)\right\rangle \right|_{\mathbf{x}=\mathbf{x}'}\; .\label{CurrentDef}\end{equation}

As mentioned, the quartic term in the thermodynamical potential will be
treated in the Hartree approximation,\cite{Ullah91,Penev0} by replacing
the cubic term $b\left|\psi _{l}\right|^{2}\psi _{l}$ in Eq. (\ref{EQini})
with $b\left\langle \left|\psi _{l}\right|^{2}\right\rangle \psi _{l}$.
This results in a linear problem with a modified (renormalized) GL potential
$\widetilde{a}=a+b\left\langle \left|\psi _{l}\right|^{2}\right\rangle $,
and a renormalized reduced temperature \begin{equation}
\widetilde{\varepsilon }=\varepsilon +\frac{b}{a_{0}}\left\langle \left|\psi _{l}\right|^{2}\right\rangle \; .\label{RenormEps}\end{equation}

We shall further introduce the Fourier transform with respect to the in-plane
coordinate $y$, the layer index $l$, and time $t$, respectively, and
also the Landau level (LL) representation with respect to the $x$-dependence,
through the relation:\begin{align}
\psi _{l}(x,y,t) & =\int \frac{dk}{2\pi }\int _{-\pi /s}^{\pi /s}\frac{dq}{2\pi }\int \frac{d\omega }{2\pi }\sum _{n\geq 0}\psi _{q}(n,k,\omega )\label{Fourier}\\
 & \cdot e^{-iky}e^{-iqns}e^{-i\omega t}u_{n}\left(x+\frac{\hbar k}{e_{0}B}\right)\, ,\nonumber
\end{align}
where the functions $u_{n}\left(x\right)$ with $n\in \mathbb{N}$ build
the orthonormal eigenfunction system of the harmonic oscillator hamiltonian,
so that $\left(-\hbar ^{2}\partial _{x}^{2}+e_{0}^{2}B^{2}x^{2}\right)u_{n}\left(x\right)=\hbar e_{0}B\left(2n+1\right)u_{n}\left(x\right)$.

The equation (\ref{EQini}) can be put further, after applying the expansion
(\ref{Fourier}), in the matrix form:\begin{equation}
\sum _{n'}\mathbf{M}_{nn'}\psi _{q}(n',k,\omega )=\zeta _{q}(n,k,\omega )\label{MatrixEq}\end{equation}
 where the symmetrical tridiagonal matrix $\mathbf{M}$ has the non-zero
elements:\begin{eqnarray}
 &  & \mathbf{M}_{00}=i\, \Gamma _{0}^{-1}\left(\frac{Ek}{B}-\omega \right)+\widetilde{a}+\frac{\hbar e_{0}B}{2m}+\frac{\hbar ^{2}(1-\cos qs)}{m_{c}s^{2}}\, ;\nonumber \\
 &  & \mathbf{M}_{nn}=\mathbf{M}_{00}+\frac{\hbar e_{0}B}{m}n\, ;\label{MatrixElem}\\
 &  & \mathbf{M}_{n+1,n}=\mathbf{M}_{n,n+1}=-i\, \Gamma _{0}^{-1}E\sqrt{\frac{e_{0}}{2\hbar B}}\, \sqrt{n+1}\, \, ,\nonumber
\end{eqnarray}
 and where the new noise terms $\zeta _{q}\left(n,k,\omega \right)$, corresponding
to the expansion rule (\ref{Fourier}), are delta-correlated such as $\left\langle \zeta _{q}\left(n,k,\omega \right)\zeta _{q'}^{*}\left(n',k',\omega '\right)\right\rangle =2\Gamma _{0}^{-1}k_{B}T(2\pi )^{3}\delta (k-k')\delta (q-q')\delta (\omega -\omega ')\delta _{nn'}$.

By solving Eq. (\ref{MatrixEq}), and taking into account the expansion
form (\ref{Fourier}), one obtains further the correlation function of
the order parameter:\begin{eqnarray}
 &  & \left\langle \psi _{l}\left(x,y,t\right)\psi _{l}^{*}\left(x',y,t\right)\right\rangle =2\Gamma _{0}^{-1}k_{B}T\label{PsiCorr1}\\
 &  & \cdot \int \frac{dk}{2\pi }\int \frac{d\omega }{2\pi }\int \frac{dq}{2\pi }\sum _{n}\sum _{n'}u_{n}\left(x+\frac{\hbar k}{e_{0}B}\right)\nonumber \\
 &  & \cdot u_{n'}\left(x'+\frac{\hbar k}{e_{0}B}\right)\left(\mathbf{M}^{*}\cdot \mathbf{M}\right)_{nn'}^{-1}\left(q,k,\omega \right)\, ,\nonumber
\end{eqnarray}
 where the expression $\left(\mathbf{M}^{*}\cdot \mathbf{M}\right)_{nn'}^{-1}$
is to be understood as the element of the inverted matrix.

It is hereafter more convenient to rescale the integration variables to
the new ones\begin{equation}
\omega '=\frac{\Gamma _{0}^{-1}}{a_{0}}\left(\omega -\frac{Ek}{B}\right);\quad q'=qs;\quad k'=\frac{\hbar }{e_{0}B}k\, ,\label{NewVariables}\end{equation}
 and to introduce the reduced field magnitudes\begin{equation}
h=\frac{B}{B_{c2}(0)}=\frac{\hbar e_{0}B}{2ma_{0}}\quad \text {and}\quad E'=\frac{Ee_{0}\xi _{0}\Gamma _{0}^{-1}}{4\sqrt{3}a_{0}\hbar }=\frac{E}{E_{0}}\; ,\label{ReducedFields}\end{equation}
 where $E_{0}=16\sqrt{3}k_{B}T\, /\, \pi e\xi _{0}$ is defined as in Refs.
\onlinecite{Varlamov92} and \onlinecite
{Mishonov02}. One can thus use further instead of the matrix $\mathbf{M}$ the following
matrix $\mathbf{M}'=a_{0}^{-1}\mathbf{M}$ having no physical dimension,
so that

\begin{eqnarray}
\mathbf{M}'_{00} & = & -i\omega '+\widetilde{\varepsilon }+\frac{r}{2}\left(1-\cos q'\right)+h\, ;\nonumber \\
\mathbf{M}'_{nn} & = & \mathbf{M}'_{00}+2hn;\label{NewMatrixElem}\\
\mathbf{M}'_{n+1,n} & = & \mathbf{M}'_{n,n+1}=-i\, 2\sqrt{6}\frac{E'}{\sqrt{h}}\, \sqrt{n+1}\nonumber \\
r & = & 2\hbar ^{2}/a_{0}m_{c}s^{2}=\left(2\xi _{0c}/s\right)^{2}\, .\nonumber
\end{eqnarray}
After writing Eq. (\ref{PsiCorr1}) in the newly introduced variables and
performing the integral over $k'$, we are able to compute the fluctuation
Cooper pair density $\left\langle \left|\psi \right|^{2}\right\rangle $
and write the self-consistent equation (\ref{RenormEps}) for the renormalized
reduced temperature parameter $\widetilde{\varepsilon }$ under the form:\begin{equation}
\widetilde{\varepsilon }=\ln \frac{T}{T_{0}}+gT\, 4h\int _{-\infty }^{\infty }\frac{d\omega '}{2\pi }\int _{-\pi }^{\pi }\frac{dq'}{2\pi }\sum _{n}\mathbf{Q}_{nn}\left(q',\omega '\right)\, ,\label{self-const-eq}\end{equation}
where we have introduced the hermitian matrix $\mathbf{Q=}\left(\mathbf{M}'^{*}\cdot \mathbf{M}'\right)^{-1}$
and the parameter $g=2\mu _{0}\kappa ^{2}e^{2}\xi _{0}^{2}k_{B}/\left(\pi \hbar ^{2}s\right)$.
The expression of the quartic term coefficient\cite{Ullah91} $b=\mu _{0}\kappa ^{2}e_{0}^{2}\hbar ^{2}/2m^{2}$
was also taken into account, with $\kappa $ being the Ginzburg-Landau
parameter $\kappa =\lambda _{0}/\xi _{0}$. It must be noticed that the
equation for the renormalized reduced temperature, $\widetilde{\varepsilon }$,
is highly nonlinear, since the parameter $\widetilde{\varepsilon }$ enters
the $\mathbf{M}'_{00}$ expression (\ref{NewMatrixElem}) and therefore
the $\mathbf{Q}$ matrix elements.

Analogously, starting from the correlation function (\ref{PsiCorr1}) written
in the new variables from Eqs. (\ref{NewVariables}) and (\ref{ReducedFields}),
we can eventually find a simpler form for the fluctuation conductivity:\begin{align}
\sigma  & =\frac{e^{2}}{4\hbar s}\frac{2h^{3/2}}{E'\sqrt{6}}\int _{-\infty }^{\infty }\frac{d\omega '}{2\pi }\int _{-\pi }^{\pi }\frac{dq'}{2\pi }\sum _{n}\sqrt{n+1}\: \mathrm{Im}\left(\mathbf{Q}_{n+1,n}\right)\: .\label{sigma1}
\end{align}
where the recursive properties of the $u_{n}\left(x\right)$functions and
the $\mathbf{Q}$ matrix hermiticity were also used.

In both expressions Eqs. (\ref{self-const-eq}) and (\ref{sigma1}), the
sum over the Landau levels must be cut-off at some maximum index $N_{c}$,
reflecting the inherent UV divergence of the Ginzburg-Landau theory. The
classical\cite{Schmid69,Penev0} procedure is to suppress the short wavelength
fluctuating modes through a \emph{momentum} (or, equivalently, \emph{kinetic
energy}) \emph{cut-off} condition, which, in terms of the Landau level
representation writes\cite{Ullah91,Penev0} $\left(\hbar e_{0}B/m\right)\left(n+\frac{1}{2}\right)\leq ca_{0}=c\hbar ^{2}/2m\xi _{0}^{2}$,
where $c$ is an adimensional cut-off parameter of the order of unity.
A \emph{total energy cut-off} was also proposed,\cite{Carballeira01} which
eliminates the most energetic fluctuations and not only those with short
wavelengths, and whose physical meaning was recently shown to follow from
the uncertainty principle,\cite{Vidal02} which imposes a limit to the
confinement of the superconducting wave function. However, in the critical
fluctuation region the two cut-off prescriptions almost coincide quantitatively,
due to the low reduced-temperature $\varepsilon $ with respect to $c$,
so that we shall apply for simplicity the cut-off procedure in its classical
form. In terms of the reduced magnetic field $h$, it writes thus $h\left(N_{c}+\frac{1}{2}\right)=c/2$.
In this way, the matrices $\mathbf{M}$, $\mathbf{M}'$ and $\mathbf{Q}$
are truncated at $N_{c}+1$ lines and columns, and the formulae (\ref{self-const-eq})
and (\ref{sigma1}) can be computed numerically.

The renormalization procedure consists thus in using the reduced temperature
parameter $\widetilde{\varepsilon }$, renormalized by solving Eq. (\ref{self-const-eq}),
in the conductivity expression (\ref{sigma1}). This procedure causes the
critical temperature to shift downwards with respect to the bare mean-field
transition temperature $T_{0}$. In analogy with the Gaussian fluctuation
case, we shall adopt as definition for the critical temperature $T_{c}(E,B)$
the vanishing of the reduced temperature, $\widetilde{\varepsilon }=0$.
In practice, one knows experimentally the actual critical temperature $T_{c}(0,0)\equiv T_{c0}$
measured at very low electrical field and with zero magnetic field, so
that one would have to consider Eq. (\ref{self-const-eq}) in the zero-fields
limit. The relationship between $T_{c0}$ and $T_{0}$ has been however
already found in Ref. \onlinecite{PuicaLangE} and writes $T_{0}=T_{c0}\left[\sqrt{c/r}+\sqrt{1+(c/r)}\right]^{2gT_{c0}}$.
Now, having $T_{0}$ one can use Eq. (\ref{self-const-eq}) for any temperature
$T$ and fields $E$ and $B$ in order to find the actual renormalized
$\widetilde{\varepsilon }(T,E,B)$, and further the conductivity $\sigma (T,E,B)$
from Eq. (\ref{sigma1}).

In order to illustrate the main features of our model, we take as example
a common material, like the optimally doped YBa$_{2}$Cu$_{3}$O$_{6+x}$.
Typical characteristic parameters are then: $s=1.17$ nm for the interlayer
distance, $\xi _{0}=1.2$ nm and $\xi _{0c}=0.14$ nm for the zero-temperature
in-plane and out-of-plane coherence lengths, respectively, $\kappa =70$
for the Ginzburg-Landau parameter and $T_{c0}=92$ K for the critical temperature
under very small electric field and no magnetic field. We also choose for
convenience a linear temperature extrapolation for the normal state resistivity
which vanishes at $T=0$, and has a typical value $\rho _{N}=84\, \mu \Omega $cm
at $T=200$ K.

\begin{figure*}
\includegraphics[  bb=33bp 35bp 853bp 330bp,
  clip,
  width=18cm]{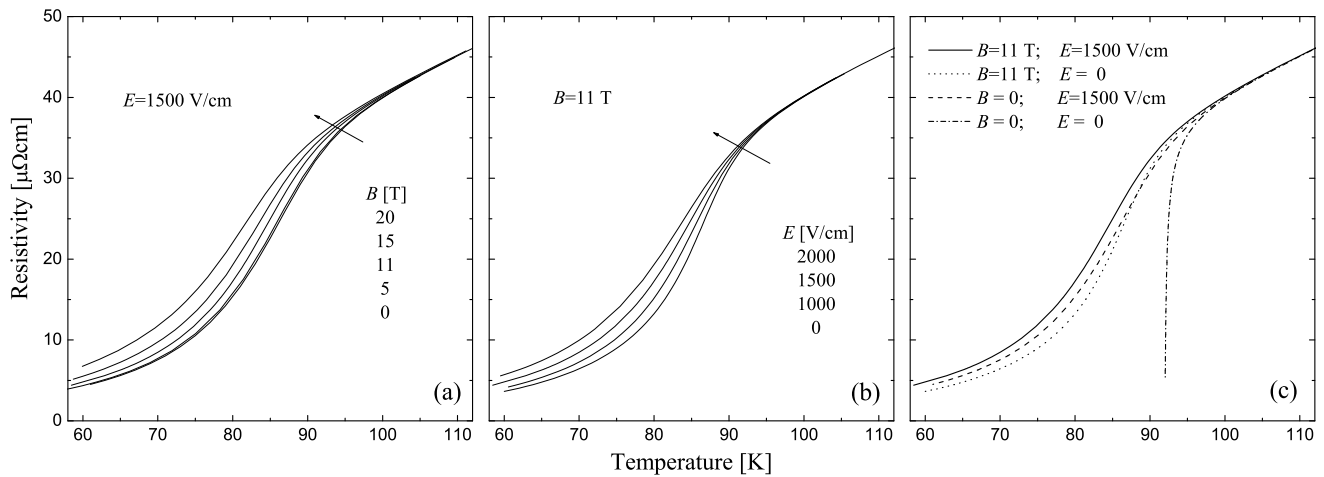}

\caption{Resistivity as a function of temperature for different combinations of
the perpendicularly applied magnetic field and the in-plane electric field.
The material parameters are given in the text. The arrows indicates the
increasing direction of the relevant variable field. (a) Fixed electric
field and variable magnetic field; the zero magnetic field curve is computed
according to the model from Ref. \onlinecite{PuicaLangE}. (b) Variable
electric field at constant magnetic field; the zero electric field curve
(linear response limit) is computed according to the UD model. The UV cut-off
parameter $c=1$ in our model corresponds with limiting the sum on the
Landau levels at the index $1/2h$ in the UD model. (c) Effects produced
by the individually (dash and dot curves) and the simultaneously (solid
curve) applied strong electric and magnetic fields; the resistivity curve
in the absence of applied fields (dash-dot curve) corresponds to the LD
model taken in the Hartree approximation and with considering the UV cut-off.\label{AllFigs}}
\end{figure*}

In Fig. \ref{AllFigs}a the resistivity curves computed according to Eqs.
(\ref{self-const-eq}) and (\ref{sigma1}) are shown for different magnitudes
of the magnetic field, at a fixed value of a strong electric field. The
zero magnetic field curve is however computed within the model presented
in Ref. \onlinecite{PuicaLangE}, that treats the non-ohmic fluctuation
conductivity in arbitrarily strong electric field and zero magnetic field,
in the Hartree approximation and with consideration of the UV cut-off.
We can notice that the numerical results obtained for finite magnetic fields
tend to approach the curve computed according to Ref. \onlinecite{PuicaLangE}
while $B$ decreases towards zero.

Figure \ref{AllFigs}b shows the complementary case of a fixed magnetic
field and different magnitudes of the in-plane electric field. The fact
that the curves approach the result from the UD model in the vanishing
electric field limit can be easily proven also analytically. Equation (\ref{self-const-eq})
can be directly written in the $E=0$ limit, and since in this case the
$\mathbf{M}$-matrix becomes diagonal, the inverted matrix $\mathbf{Q=}\left(\mathbf{M}'^{*}\mathbf{M}'\right)^{-1}$
can be trivially computed as having the elements $\mathbf{Q}_{nn}^{(0)}=\left\{ \omega '^{2}+\left[\widetilde{\varepsilon }+r\left(1-\cos q'\right)/2+h\left(2n+1\right)\right]^{2}\right\} ^{-1}$,
so that the two integrals over $\omega '$ and $q'$ can be immediately
solved and yield $\left.\widetilde{\varepsilon }\right|_{E=0}=\ln \frac{T}{T_{0}}+gT\cdot 2h\sum _{n=0}^{N_{c}}D_{n}\, $,
with $D_{n}=\left[\left(\widetilde{\varepsilon }+h+2nh\right)\left(\widetilde{\varepsilon }+h+2nh+r\right)\right]^{-1/2}$.
This is coincident with the analogous equation found by UD for the renormalized
reduced temperature in the linear response approximation.

Equation (\ref{sigma1}) can be also processed in the vanishing electric
field limit by expanding the $\mathbf{Q}$-matrix up to the linear term
in $E'$, so that one obtains $\mathrm{Im}\mathbf{Q}_{n+1,n}^{(1)}=4\sqrt{6}E'\sqrt{h}\sqrt{n+1}\, \mathbf{Q}_{nn}^{(0)}\mathbf{Q}_{n+1,n+1}^{(0)}$,
and after solving the integrals over $\omega '$ and $q'$, the result
becomes $\left.\sigma \right|_{E\rightarrow 0}=\left(e^{2}/4\hbar s\right)\sum _{n=0}^{N_{c}}\left(n+1\right)\left(D_{n}-2D_{n+\frac{1}{2}}+D_{n+1}\right)$.
This latter expression matches thus the linear response approximation for
the in-plane conductivity found by UD, as well as the corresponding formula
in the paper by Ikeda \emph{et al.}\cite{Ikeda91a}

Figure \ref{AllFigs}c shows a further illustration of the effect obtained
by the simultaneous application of the magnetic and electric fields, as
compared to the individual effects produced by each field alone. The zero-fields
curve is computed according to the LD model with the UV cut-off included,\cite{Carballeira01,PuicaLangE}
namely $\left(16\hbar s/e^{2}\right)\left.\sigma (\widetilde{\varepsilon })\right|_{E,B=0}=\left[\widetilde{\varepsilon }\left(\widetilde{\varepsilon }+r\right)\right]^{-1/2}-\left[\left(\widetilde{\varepsilon }+c\right)\left(\widetilde{\varepsilon }+c+r\right)\right]^{-1/2}-c(c+\widetilde{\varepsilon }+r/2)\left[(c+\widetilde{\varepsilon }+r)(c+\widetilde{\varepsilon })\right]^{-3/2}$,
where the renormalized reduced temperature parameter $\widetilde{\varepsilon }$
is given by the self-consistent Hartree approximation\cite{Penev0,PuicaLangE}
$\left.\widetilde{\varepsilon }\right|_{E,B=0}=\ln \left(T/T_{0}\right)+2gT\, \ln \left[\left(\sqrt{\widetilde{\varepsilon }+c}+\sqrt{\widetilde{\varepsilon }+c+r}\right)/\left(\sqrt{\widetilde{\varepsilon }}+\sqrt{\widetilde{\varepsilon }+r}\right)\right]$.

The values for the electric and magnetic fields in Fig. \ref{AllFigs}c
were chosen such that their individual effect be almost similar. One can
notice that the simultaneous application of the two fields brings however
only a slight enhancement of the superconducting fluctuation suppression.
Nevertheless, the non-ohmic behaviour of the fluctuation conductivity at
high electric fields still remains quantitatively important for commonly
used HTSC also in the presence of strong magnetic fields, so that experimental
investigations could be able to discern between the applicability of this
model in competition with the linear response approximation.

In summary, we have treated the critical fluctuation conductivity for a
layered superconductor in perpendicular magnetic field, in the frame of
the self-consistent Hartree approximation, for an arbitrary in-plane electric
field value. The main results are the formulae (\ref{sigma1}) for the
fluctuation conductivity, and (\ref{self-const-eq}) for the renormalized
reduced-temperature parameter. In the two limit cases, namely for ($E=0$,
$B>0$) and for ($E>0$, $B=0$), the corresponding solutions were found,
analytically and respectively numerically, to reduce to previous results
of existing theories. Qualitatively, the simultaneous application of the
two fields results in a slight additional suppression of the superconducting
fluctuation, compared to the case when the fields are individually applied.

This work was supported by the Austrian Fonds zur F\"{o}rderung der wissenschaftlichen
Forschung.

\bibliographystyle{APSREV}
\bibliography{PuicaLang2}

\end{document}